\documentstyle[12pt]{article}
\oddsidemargin--0mm
\begin{document}
\newcommand{\la}{\langle}
\newcommand{\ra}{\rangle}
\newcommand{\pl}{\partial}
\newcommand{\be}{\begin{equation}}
\newcommand{\ee}{\end{equation}}
\newcommand{\ba}{\begin{eqnarray}}
\newcommand{\ea}{\end{eqnarray}}
\def\R{\relax{\rm I\kern-.18em R}}
\def\1{\relax{\rm 1\kern-.27em I}}
\newcommand{\Z}{Z\!\!\! Z}
\newcommand{\ph}{PS_{ph}}
\begin{titlepage}
\begin{center}
{\huge A Transition Amplitude on the Gauge Orbit Space}$ ^{ ^{ ^*}}$

\vskip 1cm
{\large\bf Sergey V. SHABANOV}$ ^{ ^{ ^{**}}}$

\vskip 1cm
{\em Service de Physique Theorique de Saclay,\\
CEA-Saclay, Gif-sur-Yvette Cedex, F-91191,\\
 France}
\end{center}

\vskip 2cm
\begin{abstract}
A general procedure for deriving the path integral representation
of a transition amplitude on the gauge orbit space having a non-trivial
topology is proposed. The path integral formula appears to be modified
by including trajectories reflected from the physical configuration space
boundary into the sum over paths.
A solution of the Gribov problem of gauge fixing
ambiguities is given in the framework of the path integral modified.
\end{abstract}

\vskip 4cm
\noindent
Submitted to "{\em The Proceedings of the II International Workshop
on Constrained Systems}", Montepulchiano, Italy, June, 1993.

\vskip 3cm
\noindent
\underline{\hspace*{8cm}}

\noindent
$ ^*$Work supported by the MRT grant of the government of France.

\noindent
$ ^{**}$On leave from: {\em Laboratory of Theoretical Physics,
Joint Institute for Nuclear Research, P.O.Box 79, Moscow, Russia}

\noindent
e-mail address: {\bf shabanov@amoco.saclay.cea.fr}
\end{titlepage}

It is well-known that quantum dynamics strongly depends on
a configuration space topology. For example, a spectrum of
a free particle moving on a line is continuous, while
the same system on a circle has a discrete spectrum.
Therefore one can expect that the path integral (PI) representation
of a transition amplitude \cite{fey},\cite{kla} depends on the
configuration space topology. Wave functions of systems with
topologically non-trivial configuration spaces usually obey
some boundary conditions. For instance, wave functions of a
particle on a circle satisfy periodic boundary conditions,
while wave functions of a particle in a box must vanish at the
box walls. Thus, the problem of constructing the PI formalism
for such theories is to take into account boundary conditions
which contain all information about the configuration space topology.

Naively, one would think that it is sufficient to restrict the
integration domain in the path integral to obtain a transition
amplitude for a particle in a box or on a circle. However, we know
from the operator formalism that transition amplitudes for two
systems with the same Hamiltonian and the same volume of the configuration
space might be different if they have different boundary conditions.
A trivial example is a particle moving in a one-dimensional interval
of size $L$. Zero boundary conditions imply infinite walls
attached at the interval ends (a particle in a box),
while the periodic ones are for a particle on a circle.
The corresponding transition amplitudes
are different \cite{pau}-\cite{kle}. Thus, a restriction
of the integration domain does not lead to a correct solution of the
problem.

For simplest systems, like a particle in a box or on a circle, a
correct PI formula for a transition amplitude is obtained by including
trajectories reflected from the box walls or trajectories with
all possible winding numbers, respectively, into the Feynman sum
over paths \cite{pau}- \cite{kle} rather then by restricting the
integration domain. The method of reflected trajectories can be
successfully applied to systems with more complicated configuration spaces
\cite{q} (see for a review \cite{book} and references therein). This
structure of PI is established even for such unusual systems as the
$q$-deformed ones \cite{qd}.

Gauge theories (or the first class-constrained systems \cite{dir})
give another example of this kind. Their main feature is the existence
of unphysical variables whose evolution is not determined by equations
of motion and appears to be completely arbitrary. This arbitrariness
occurs through the gauge invariance of a corresponding Lagrangian
\cite{dir}. In the whole (total) configuration (or phase) space,
there are points related to each other by gauge transformations depending
on arbitrary functions of time. These points form orbits of the
gauge group. Any motion of a system along a gauge orbit does not lead to
any change of a physical state of a system \cite{dir}.
Two physically distinct states correspond to different gauge orbits.
Therefore, to construct a physical configuration space $CS_{ph}$
(each its point corresponds, by definition,
to just one physical state of a system), one should stick together
all points of each gauge orbit. In the mathematical language,
the physical configuration space is the quotient of the total configuration
space (CS) by the gauge group $G$,
\be
CS_{ph}=CS/G\ .
\ee
The gauge orbit space (1) might differ from an ordinary Euclidean
space \cite{lv}-\cite{ufn} and possess a non-trivial topology even
if the total CS is assumed to be Euclidean. For example, it turns
out to be compact for $QCD_2$ \cite{pisa},\cite{het} and coincides
with the Weyl cell (being a polyhedron) for a compact gauge group
\cite{pisa}.

In the present paper, we shall give a general recipe for deriving
the PI representation of a transition amplitude on the gauge orbit space (1).

Consider a quantum theory determined by the Schroedinger equation
\be
\left(-\frac12\langle\frac{\pl}{\pl x},\frac{\pl}{\pl x}\rangle
+ V(x)\right)
\psi_E = E\psi_E\ .
\ee
The eigen-functions $\psi_E$ are normalized by the condition
\be
\int\limits_{\R^N}dx\psi_E^*(x)\psi_{E'}(x)= \delta_{EE'}\ .
\ee
We assume $x$ to realize a linear representation of a compact group $G$
so that its action is given by a linear operator
$x\rightarrow T(\omega)x$, and $V(Tx)=V(x)$;
$\langle x,y\rangle = \sum_1^Nx_iy_i = \langle Tx,Ty\rangle$ is an invariant
scalar product in the representation space that is isomorphic to $\R^N$. The
theory turns into the gauge one if we require that physical states are
annihilated by operators generating $G$-transformations of $x$,
$\hat{\sigma}_a\Phi(x)=0$. These conditions determine a physical
subspace in the Hilbert space. By definition, we have $\exp(\omega_a
\hat{\sigma}_a)\psi(x) = \psi(T(\omega)x) $.
Therefore, the physical states are $G$-invariant
\be
\Phi(T(\omega)x)= \Phi(x)\ ,
\ee
where $\omega$ runs over the group manifold.

Let a number of physical degrees of freedom in the system is
equal to $M$, then a number of independent constraints is $N-M$.
Suppose we would like to span the physical configuration space
$K\sim \R^N/G$ by coordinates ranging a gauge condition surface
$F(x)=0$. We assume the gauge condition to be complete, meaning that
there is no unphysical degree of freedom left. Let $u\in \R^M$ be
a parameter of the gauge condition surface $x=f(u)$ such that
$F(f(u))$ identically vanishes for all $u\in \R^M$.
The gauge condition surface must cross each gauge orbit at least once
so that $u$ could serve as a coordinate spanning $K$. In other
words, any particular configuration $x$ can be transferred onto
the gauge condition surface by a gauge transformation, i.e.
$x$ can be represented as follows
\be
x=x(\theta,u)=T(\theta)f(u)\ .
\ee

Physical states (4) are independent of the variables $\theta$ spanning
the gauge group manifold. Therefore, a projection of  the Hamiltonian
operator on the physical subspace can be done by introducing the
curvilinear coordinates (5) in the Laplace operator entering into (2)
and by dropping all terms containing derivatives with respect to $\theta$
in it. Provided each gauge orbit intersects a gauge condition surface
just once, we would have a quantum theory on a curved configuration
space which is topologically equivalent to $\R^M$. A construction of
the PI representation for a transition amplitude in this case does not
give rise to any serious problem, and is known \cite{kle},
\cite{book}. Troubles
appear when a gauge condition does not completely fix a gauge arbitrariness,
meaning that there are configurations on the gauge condition surface which
are connected with each other  by gauge transformations. The residual
gauge arbitrariness cannot decrease a number of physical degrees of
freedom, but it does reduce their configuration space.

The existence of residual gauge transformations means that the
gauge condition surface intersects each gauge orbit not only once.
All the intersection points are, obviously, gauge copies of one of them
\cite{gri}.
The latter may occur due to two reason. The fist one is that a gauge
condition looks preferable because of physical reasons in spite of its
ambiguity. Another reason is hidden in a mathematical structure of
gauge systems. One cannot always fix a gauge without any ambiguity due to
a non-trivial topology of the gauge orbit space. Yang-Mills fields give
an example of such a kind \cite{gri},\cite{sin}.

As has been pointed out above,
a configuration space topology is taken into
account by boundary conditions imposed on wave functions. Therefore, to
construct the PI representation of a transition amplitude on the gauge
orbit space, one should first find boundary conditions for physical
states appearing upon projecting the Hamiltonian operator on the orbit
space, and then to  derive PI corresponding to them. To go over our
program, we shall introduce the curvilinear coordinates (5) to remove
unphysical degrees of freedom. The boundary conditions determining
the gauge orbit space topology will be shown to result from analytical
properties of physical wave functions.

The metric tensor in the new coordinates reads
\be
\langle dx,dx \rangle = \langle df, df \rangle + 2\langle df,d\theta f
\rangle + \langle d\theta f, d\theta f\rangle \equiv g_{AB}dy^Ady^B\ ,
\ee
where we have used the $G$-invariance of the bilinear form $\la,\ra$ and
put $d\theta =T^{-1}dT$ and $dy^1\equiv du,\ dy^2\equiv
d\theta$. Therefore,
\be
\int\limits_{\R^N}dx = \int\limits_{G}\wedge d\theta
\int\limits_K d^M u\mu(u)\ ;
\ee
here $\mu(u)= (\det g_{AB})^{1/2},\ K$ is a subdomain in $\R^M$ such
that the mapping (5), $K\otimes G\rightarrow \R^N$, is one-to-one.
To determine $K$, one should find transformations $\theta,u
\rightarrow \hat{s}\theta, \hat{s}u,\ \hat{s}\in \tilde{S}_F$, which
leave $x$ untouched, $x(\hat{s}\theta, \hat{s}u)=x(\theta,u)$.
 Obviously, $\tilde{S}_F=T_e\times S_F$ where $T_e$ is a
group of translations of $\theta$ through periods of the group manifold
$G$ and $\hat{s}u = u,\ \hat{s}\in T_e$,
while the set $S_F$ is obtained by solving the following equation
\be
F(T_sf(u))=0
\ee
with respect to a gauge transformation operator $T_s$. Indeed, assuming
Eq.(8) to have non-trivial solutions (the trivial one $T_s=1$ always
exists by the definition of $f(u)$) we observe that all points
$T_sf$ belong to the gauge condition surface and, hence, $T_sf(u)=
f(u_s), \ u_s=u_s(u)$. Consider transformations of $\theta$ generated
by the group shift $T(\theta)\rightarrow T(\theta)T^{-1}_s=T(\theta_s),\
\theta_s=\theta_s(\theta,u)$. Setting $\hat{s}u=u_s$ and $\hat{s}
\theta= \theta_s$ we see that the transformations $\hat{s}\in S_F$
leave $x=x(\theta,u)$ untouched. To avoid a "double" counting in the
scalar product integral (7) (cf. (3)),
one has to restrict the integration domain for $u$ to the
quotient $\R^M/S_F=K$. The modular domain $K$ can also be determined
by the requirement that a part of the gauge condition surface $x=f(u),\
u\in K\subset \R^M$ has just one common point with any gauge orbit.

A choice of the modular domain parametrization is not unique. Suppose
we find a part of a given gauge condition surface such that each
gauge orbit is represented by one point on it, i.e. we identify
$K$ with a concrete subregion in $\R^M$. Apparently, any subregion
among $\hat{s}K=K_s,\ \hat{s}\in S_F$, can serve as the integration
region in (7).
Having chosen a concrete parametrization (coordinates)
of $K$, we fix a representation of $S_F$ by functions
$\hat{s}u=u_s(u),\ u\in K,\ u_s\in K_s,\ K_s\cap K_{s'}=\emptyset $ for
any $\hat{s}\neq\hat{s}'$ and $\R^M= \cup_s K_s$ up to a set of zero
measure being a unification of the boundaries $\pl K_s$.

The functions $u_s(u)$ might not have an analytical continuation to the
whole covering space $u\in \R^M$. Let $u_s$ and $u_{s'}$ have
an analytical continuation to $\R^M$. The latter would mean that their
composition $u_s\circ u_{s'}$ exists and must determine an operator
$T_{ss'}$ obeying (8) since there should be $T_{ss'}=T_sT_{s'}$.
However, a composition of two solutions to (8) does not always give
a new solution because $T_s=T_s(u)$ and $T_{s'}= T_{s'}(u)$ are
functions of $u$, i.e. in a general case $F(T_s(u)T_{s}(u)f(u)) =
F(T_s(u)f(u_{s'})) \neq 0 $ since $u_{s'}\neq u$, whereas
$F(T_s(u)f(u)) =0$.
Herein we restrict ourselves just by pointing out
this fact and by illustrating it with a simple example (see it
below). We shall assume that the functions $u_s$ are analytical
only on $K$, while $u_s^{-1}$ maps $K_s$ on $K$, which is sufficient
for what follows. So, we admit the set of transformations $S_F$ to be
not a group since a composition of its elements might not be uniquely
defined. Notice also that residual gauge transformations may not form a
group for Yang-Mills theories \cite{sol}.

We define
an orientation of $K_s$ so that for all $\hat{s}\in S_F$,
$\int_{K_s}du\phi\geq 0$ if $\phi\geq 0$, which provides the following
rules
\ba
\int\limits_{\R^M}d^Mu &=& \sum\limits_{S_F}\int\limits_{K_s}d^Mu\ ,\\
\int\limits_Kd^Mu |J_s(u)|&=& \int\limits_{K_s}d^Mu\ ,
\ea
where $J_s(u) = Du_s/Du$ is the Jacobian, the absolute value of
$J_s$ has been inserted into the left-hand side of (10) for
preserving the positive orientation of the integration domain.
We assume the Jacobian $\mu$ in (7) to be positive on the domain
$K$ chosen, otherwise one should take its absolute value in accordance
with our orientation rules.

{\em Remark}. A number of elements in $S_F$ can depend on $u$.
We define a region $\R^M_\alpha\subseteq \R^M $ such that
$S_F=S_\alpha$ has a fixed number of elements for all $u\in
\R^M_\alpha$. Then $K=\cup_\alpha K_\alpha,\ K_\alpha
=\R^M_\alpha/S_\alpha,\ \R^M= \cup_\alpha\R^M_\alpha$. The
sum in (9) implies $\sum_{S_F}=\sum_\alpha\sum_{S_\alpha}$
and $K_s$ in (9-10) carries an additional suffix
$\alpha$. In what follows we shall omit it and use the simplified
notations (9-10) to avoid complications of formulas. The suffix
$\alpha$ can be easily restored  by means of the rule
proposed above.

{\em Example}. Let the whole configuration space be a plane $\R^2$ and the
gauge group be an $SO(2)$-rotation of $\R^2$. Gauge orbits are
concentric circles. Any gauge condition $F({\bf x}) =0,\ {\bf x}\in \R^2$,
determine a curve ${\bf x}={\bf f}(u),\ u\in\R$, which goes
through the origin to infinity (to provide crossing each orbit
at least once). So, $T(\theta)=\exp(i\theta\sigma_2),\ \sigma_2$
the Pauli matrix, and $x,f\rightarrow {\bf x,f}\in \R^2$ in (5).
Set $f_1= -u_0,\ f_2=-\gamma(2u_0 + u)$ for $u< - u_0$ and $f_1= u,\
f_2=\gamma u$ for $u> -u_0$ where
$\gamma $ and $u_0$ are positive constants. The
curve $x_{1,2}=f_{1,2}(u)$
 touches circles (gauge orbits) of radii $r= u_0$ and
$r=u_0\gamma_0,\ \gamma_0=\sqrt{1+\gamma^2}$. It intersects twice all
circles with radii $r< u_0$ and $r> u_0\gamma_0$, whereas any circle
with a radius from the interval $r\in (u_0,u_0\gamma_0)$ has four
common
points with the gauge condition curve. Therefore, $S_F$ has one
nontrivial element for $u\in \R_1\cup\R_3,\ \R_1= (-u_0/\gamma_0,
u_0/\gamma_0),\ \R_3= (-\infty,-3u_0)\cup (u_0,\infty)$ and three
nontrivial elements for $u\in \R_2= (-3u_0,-u_0/\gamma_0)\cup
(u_0/\gamma_0, u_0)$.
Since points ${\bf f}(u_s)$ and ${\bf f}(u)$ belong to the same
circle (gauge orbit), the functions $u_s$
have to obey the following equation
\be
{\bf f}^2(u_s)= {\bf f}^2(u)\ .
\ee
Denoting $S_F = S_\alpha$ for $u\in \R_\alpha,\
\alpha = 1,2,3$ (see {\em Remark} above),
we have $S_1=\Z_2,\ u_s(u)= -u;\ S_2$ is determined by
the following mappings of the interval $K_2 = (u_0/\gamma_0,u_0)$
\ba
u_{s_1}(u) &=& -u\ ,\ \ \ \ \ \ \ \ \ \ \ \ \ \ \ \ \ \ \ \ \ \ \ \ \ \
\ \ \ \ \ \ \ \
u_{s_1}:\ \ K_2\rightarrow (-u_0,-u_0/\gamma_0)\ ;\\
u_{s_2}(u) &=& -2u_0 + \gamma_0(u^2-u_0^2/\gamma_0^2)^{1/2}/\gamma\ ,\ \
u_{s_2}:\ \ K_2\rightarrow (-u_0,-2u_0)\ ;\\
u_{s_3}(u)&=&-2u_0 - \gamma_0(u^2-u_0^2/\gamma_0^2)^{1/2}/\gamma\ ,\ \
u_{s_3}:\ \ K_2\rightarrow (-2u_0,-3u_0)\ ;
\ea
and for $S_3$ we get
\be
u_{s}(u)=-2u_0 - \gamma_0(u^2-u_0^2/\gamma_0^2)^{1/2}/\gamma\ :
\ \ \ (u_0,\infty)\rightarrow (-\infty,-3u_0)\ .
\ee
The functions (12-14) do not have a unique analytical continuation
to the whole domain $\R_2$ and, hence, their composition is ill-defined.
The mappings (12-14) do not form a group. Since they realize a representation
of $S_\alpha$, $S_\alpha$ is not a group.

The physical configuration space is, obviously, isomorphic to
 $K= \cup K_\alpha,\
K_\alpha= \R_\alpha/S_\alpha$, i.e. $K_\alpha$ is a fundamental domain
of $\R_\alpha$ with respect to the action of $S_F=S_\alpha$ in $\R_\alpha$,
$\R_\alpha = \cup \hat{s}K_\alpha,\ \hat{s}$ ranges over $S_\alpha$.
Upon solving (8) (or (11)) we have to choose a particular interval
as the fundamental domain. We have put $K_2 = (u_0/\gamma_0,u_0)$ in
(12-14). Another choice  would lead to another form of the functions
$u_s$ (to another representation of $S_F$ in $\R_2$). Setting, for
example, $K_2= (-2u_0,-u_0)$ we obtain from (11)
\ba
u_{s_1}(u)&=& - 4u_0-u\ ,\ \ \ \  \ \ \ \ \ \ \ \ \ \ \ \ \ \ \ \ \ \ \ \ \
u_{s_1}:\ \ K_2\rightarrow (-3u_0,-2u_0)\ ;\\
u_{s_2}(u)&=& -(u_0^2+\gamma^2(2u_0 +u)^2)^{1/2}/\gamma_0\ ,\ \
u_{s_2}:\ \ \ K_2\rightarrow (-u_0,-u_0/\gamma_0)\ ;\\
u_{s_3}(u)&=& (u_0^2 +\gamma^2(2u_0 +u)^2)^{1/2}/\gamma_0\ ,\ \ \ \!
\ \ u_{s_3}:\ \ K_2\rightarrow (u_0/\gamma_0,u_0)\ .
\ea

To find group elements $T_s(u)$ corresponding to $u_s(u)$, one
should solve the equation $T_s {\bf f}(u) = {\bf f}(u_s)$.
Setting $T_s =\exp(i\omega_s\sigma_2)$
and taking $u_s$ from (12-14) we find
\ba
\omega_{s_1}(u) &=& \pi\ ;\\
\omega_{s_2}(u) &=& \frac{3\pi}{2} - \sin^{-1}\left(
\frac{u_0}{\gamma_0u}\right) - \tan^{-1}\gamma\ ;\\
\omega_{s_3}(u) &=& \frac{\pi}{2} + \sin^{-1}\left(
\frac{u_0}{\gamma_0u}\right) - \tan^{-1}\gamma\ ,
\ea
where $u\in K_2=(u_0/\gamma_0,u_0)$. Elements of $S_{1,3}$ are
obtained analogously. It is readily seen that $\Omega_{s_1}
\Omega_{s_2}\neq \Omega_{s_3}$, etc., i.e. the elements $\Omega_s$
do not form a group. An alternative choice of $K_2$ results in a
modification of the functions (19-21).

For the curvilinear coordinates (5) we have
$ \det g_{AB} = ({\bf f}',{\bf f})^2 = \mu^2(u)$.
Set $K=\cup_\alpha K_\alpha,\
K_1=(0,u_o/\gamma_0),\ K_2=(u_0/\gamma_0, u_0),\ K_3$\
$ = (u_0,\infty)$,
i.e. $K=\R_+$, then $\int_{-\infty}^\infty du =
\sum_\alpha\int_{\R_\alpha}du$ and (9) means that the upper
integral limit is always greater than the lower one, for example,
$$
\int\limits_{\R_2}du =\left(\int\limits_{-3u_0}^{-2u_0}+
\int\limits_{-2u_0}^{-u_0} +
\int\limits_{-u_0}^{-u_0/\gamma_0}+
\int\limits_{u_0/\gamma_0}^{u_0}\right) du\ ,
$$
where the terms of the sum correspond to integrations over
$\hat{s}_3K_2,\ \hat{s}_2K_2,\ \hat{s}_1K_2$ and $K_2$,
respectively (cf. (12-14)). The following chain of equalities
is to illustrate the rule (10)
\be
\int\limits_{\hat{s}_3K_2} du_{s_3} = \int\limits_{-3u_0}^{-2u_0}
du_{s_3} = \int\limits_{u_0}^{u_0/\gamma_0} du J_{s_3} =
- \int\limits_{u_0/\gamma_0}^{u_0}du J_{s_3}= \int\limits_{K_2}
du |J_{s_3}|\ ;
\ee
the last equality results from $J_{s_3}= du_{s_3}/du < 0$
(cf. (14)).

By means of the curvilinear coordinates (5) we can naturally
 incorporate a gauge condition chosen into the Dirac operator
method \cite{dir} of quantizing first-class constrained systems.
Solutions of the equation $\hat{\sigma}_a\tilde{\Phi}(x)=0$ are
given by functions independent of $\theta$,
\be
\tilde{\Phi}(x)=\tilde{\Phi}(T(\theta)f(u))=
\tilde{\Phi}(f(u))=\Phi(u)\ ,
\ee
because $\hat{\sigma}_a$ generate shifts of $\theta$ and leave
$u$ untouched. To obtain a physical Hamiltonian, one has to write
the Laplacian in (2) via the new variables (5) and omit all
terms containing derivatives with respect to $\theta$. In so doing,
we get
\be
\hat{H}^f_{ph}\Phi_E(u)=
\left(\frac12\hat{p}_ig^{ij}_{ph}\hat{p}_j + V_q(u) +
V(f(u))\right)\Phi_E(u) = E\Phi(u)\ ;
\ee
here we have introduced hermitian momenta $\hat{p}_i = -i\mu^{-1/2}
\pl_j\circ \mu^{1/2},\ \pl_j =\pl/\pl u^j$;
the metric $g^{ij}_{ph}$ in the physical
configuration space is the $11$-component of a tensor $g^{AB}$ inverse
to $g_{AB},\ g^{AC}g_{CB}=\delta^A_B,\ g^{ij}_{ph}= (g^{11})^{ij},\
i,j=1,2,...,M$; a quantum potential
\be
V_q=\frac{1}{2\sqrt{\mu}}(\pl_ig^{ij}_{ph})\pl_j\sqrt{\mu} +
\frac{1}{2\mu}g^{ij}\pl_i\pl_j\sqrt{\mu}\
\ee
appears due to the chosen ordering of
the operators $\hat{u}^i$ and $\hat{p}_i$ in the Laplace-Beltrami
operator. Because of (23), the scalar product (3) is reduced to
\be
\int\limits_{\R^N}dx\Phi^*_E(u)\Phi_{E'}(u)\rightarrow
\int\limits_Kd^Mu \mu(u)\Phi^*_E(u)\Phi_{E'}(u)=\delta_{EE'}\ ,
\ee
where a gauge orbit volume (integral over $G$ (see (7)) has
been included into norms of physical states, which we denoted by
the arrow in (26). A construction of an operator
description of a gauge theory in a given gauge condition is completed.

Notice, in this approach the variables $u$ appear to be gauge-invariant,
they parametrize the physical configuration space $CS_{ph}=K=\R^N/G$.
Two different choices of $f(u)$ in (5) implies two different parametrizations
of $CS_{ph}$ related to each other by a change of variables
$u=u(\tilde{u})$ in (24-26). Therefore quantum theories with different
$f$'s are unitary equivalent \cite{gr}, \cite{book}.

To illustrate this statement, we consider again the simplest case
\cite{lis}
$G=SO(2),\ M=1,\ g_{ph}={\bf f}^2(u)/\mu^2(u)$, and compare descriptions
in the coordinates (5) and in the polar ones ($f_1=r, f_2=0$).
With this purpose we change variables $r=r(u)=\sqrt{{\bf f}^2(u)}$
in (24-26). For
$u\in K$ the function $r(u)$ is invertible, $u=u(r), r\in \R_+$.
Simple straightforward calculations \cite{lis} lead us to the
following equalities $\hat{H}^f_{ph}=1/2\hat{p}_r^2 + V_q(r)
+ V,\ \hat{p}_r=-i r^{-1/2}\pl_r \circ r^{1/2},\ V_q
=-(8r^2)^{-1},\ \int_K du\mu = \int_0^\infty drr$. It is nothing
but quantum mechanics of a radial motion on a plane. All theories
with different $f$'s are unitary equivalent to it and, therefore,
to each other. One should stress that the operator ordering we
obtained by applying the Dirac method plays the crucial role
in providing this unitary equivalence. Another ordering of
operators in (24) would break this property.

A few observations resulting from our consideration have to
be emphasized.

1. All regular solutions of (24) have a unique analytical
continuation  to the whole space $u\in \R^M$, and they are
$S_F$-invariant,
\be
\Phi_E(u_s(u))=\Phi_E(u)\ ,\ \ \ \ \ \  \ u\in K\ .
\ee
For a proof, we point out that any regular solution of (24) is a
projection of a regular $G$-invariant solution of (2) on $K$
determined by (23). The last equality in (23) defines the
analytical continuation of $\Phi_E(u)$ because we assume
$f(u)$ to be analytical on $\R^M$; (27) follows from
the second equality in (23) and
the definition $T_sf(u)=f(u_s)$. Another proof of (27) is to
use averaging over the group manifold. Let $\Psi_E(x)$ be a
solution to (2). Then a solution to (24) is obtained by averaging
$\Psi_E(x)$ over the group manifold. Hence,
$$
\Phi_E(u) =\frac{1}{V_G}\int\limits_Gd\mu_G(\theta)\Psi_E(T(\theta)f(u))
=\frac{1}{V_G}\int\limits_Gd\mu_G(\theta)\Psi_E(T(\theta)T_sf(u)) =
\Phi_E(u_s)\ ,
$$
where $V_G$ is a group volume, $d\mu_G$ in the right-invariant Haar measure,
$d\mu(\theta_s)= d\mu(\theta)$ with $T(\theta_s) = T(\theta)T_s^{-1}$.

The condition (27) determines
the boundary condition which should be taken into account
in the PI representation of the transition amplitude on the gauge orbit
space.

2. Any amplitude, i.e. the scalar product (26) of two $S_F$-invariant
states, is independent of a $CS_{ph}$ parametrization (of a gauge choice).
An $S_F$-invariant regular function of $u\in\R^M$ is a linear combination
of the basis states $\Phi_E(u)$. Our statement follows from the unitary
equivalence of the
theories (24-26) corresponding different parametrizations of $CS_{ph}$.

3. The physical Hamiltonian in (24) is $S_F$-invariant
\be
\hat{H}^f_{ph}(u_s) =\hat{H}^f_{ph}(u)\ ,\ \ \ \ \ \ u\in K.
\ee
Let us write the Laplace-Beltrami operator $\la\pl/\pl x,\pl/\pl x\ra
=\Delta(\theta, u)$ in the variables (5),
push all derivatives $\pl_\theta$ in it to the
right by commuting them with $\theta$ and then set $\pl_\theta=0$.
We denote the operator thus obtained $\Delta_{ph}=\Delta(\theta, u)
\vert_{\pl_\theta=0}$. For any physical state $\Phi=\Phi(u)$, we
have $\Delta(\theta,u)\Phi=\Delta_{ph}\Phi$ because $\hat{\sigma}
\sim \pl_\theta$. Due to the gauge-invariance, the Hamiltonian in (2)
commutes with the constraints, $[\hat{H},\hat{\sigma}]=0$ and,
hence, $[\Delta(\theta,u),\hat{\sigma}]=0$ (the potential $V$
is $G$-invariant). Gathering the definition of $\Delta_{ph}$ and
$G$-invariance of $\Delta$ we conclude that $\Delta_{ph}=
\Delta_{ph}(u)$ is independent of $\theta$ (otherwise we would arrive to
the contradiction $0=[\Delta,\hat{\sigma}]\Phi =
\hat{\sigma}\Delta\Phi = \hat{\sigma}\Delta_{ph}\Phi \sim
\pl_\theta\Delta_{ph}\Phi \neq 0$). Consider now the change
of variables $\theta,u\rightarrow \theta_s,u_s$. By its definition
$\Delta(\theta_s,u_s)=\Delta(\theta,u)$ and $\pl_\theta\sim
\pl_{\theta_s}$ (i.e. $\pl_\theta$ does not contain a term
proportional $\pl_{u_s}$ since $\pl u_s/\pl_\theta =0$). This
yields $\Delta_{ph}(u_s)= \Delta(\theta_s,u_s)\vert_{\pl_{\theta_s}=0}
=\Delta(\theta_s,u_s)\vert_{\pl_\theta =0} = \Delta(\theta,u)
\vert_{\pl_{\theta}=0} =\Delta_{ph}(u)$, which completes the
proof of (28).

To derive a path integral representation of the quantum theory (24-26),
we consider a slice approximation of the transition amplitude
$U_t^{ph}(u,u')=\langle u|\exp(-i\hat{H}_{ph}t)|u'\rangle$,
\be
U_t^{ph}(u,u')=\lim\limits_{\epsilon\rightarrow 0}\int\limits_K
\prod\limits_{k=0}^n\left(d^Mu_k\mu(u_k)\right)
U_\epsilon^{ph}(u,u_n)U_\epsilon^{ph}(u_n,u_{n-1})\cdots
U_\epsilon^{ph}(u_1,u')\ ,
\ee
where $(n+1)\epsilon =t$, the limit is taken so that $n\rightarrow \infty,
\ \epsilon\rightarrow 0$, while $t$ is kept fixed; the infinitesimal
evolution operator kernel reads
\be
U_\epsilon^{ph}(u,u')= [1-i\epsilon\hat{H}_{ph}(u)]
\langle u|u'\rangle + O(\epsilon^2)\ .
\ee

A naive limit in (29) gives a formal PI with a restricted integration
domain $K\subset \R^M$. A calculation of such a PI meets difficulties
because even a finite dimensional Gaussian integral cannot be explicitly
done over a part of an Euclidean space. In addition, a restriction of
the PI integration domain is meaningless for systems with boundary
conditions appearing  due to a non-trivial topology of a configuration
space like for a particle in a box or on a circle \cite{pau}-\cite{book}.
Topological properties of a configuration space
 are taken into account in PI by including additional ``reflected''
trajectories into the sum over paths \cite{q},\cite{book} rather than by
restricting the PI integration domain. Technically, a relation
between transition amplitudes $U_t$ and $U_t^{eff}$ for the same
systems (the same Hamiltonian) in a topologically non-trivial
$CS$ and in $CS=\R^M$, respectively, is established by means of
an operator $\hat{Q}$ containing all information about a CS topology,
$\hat{U}_t = \hat{U}_t^{eff}\hat{Q}$ \cite{q},\cite{book}. The same form
of PI turns out to be valid for many particular gauge models
\cite{book},\cite{weyl}-\cite{pisa}, \cite{gr}.
Bellow we shall prove this for our general case..
For deriving a PI formula we shall use
a method of an analytical continuation of the unit operator kernel
\cite{gr},\cite{book}.

The unit operator kernel $\langle u|u'\rangle$ has a natural
analytical continuation to the unphysical domain $u\in \R^M$.
Indeed, due to the $S_F$-invariance of the basis states (27)
and the scalar product form (26) we have
\be
\langle u|u'\rangle =\sum\limits_E\Phi_E(u)\Phi^*_E(u') =
\sum\limits_{S_F}\left(\mu(u)\mu(u')\right)^{-1/2}
\delta^M(u-\hat{s}u')\ ,
\ee
where $u\in \R^M,\ u'\in K$. Representing $\delta$-functions
in (31) through the Fourier integral, substituting it into (30)
and calculating  the
action of $\hat{H}_{ph}(u)$ on the unit operator kernel, we obtain
\ba
U_\epsilon^{ph}(u,u')&=&\int\limits_{\R^M}\frac{d^Mu''}
{(\mu\mu'')^{1/2}}U_\epsilon^{eff}(u,u'')Q(u'',u')\ ,\\
U_\epsilon^{eff}(u,u'')&=&\int\limits_{\R^M}\frac{d^Mp}
{(2\pi)^M}\exp\left[i\epsilon\left(p_j\frac{\Delta_j}{\epsilon}
- H^{eff}(u,p)\right)\right]\ ,\\
H^{eff}(u,p)&=&\frac12 g^{ij}_{ph}(u)p_ip_j + \frac i2
\pl_ig^{ij}_{ph}(u)p_j + V_q(u) +V\ ,\\
Q(u'',u')&=&\sum\limits_{S_F}\delta^M(u''-\hat{s}u'),\ \
\ \ u''\in \R^M\ ,\ \ \ u'\in K\ ,
\ea
where $\mu''=\mu(u'')$ and $\Delta_j=u_j-u_j''$. So, the
infinitesimal evolution operator kernel (32) has the
desired form $\hat{U}_\epsilon^{ph}=\hat{U}_\epsilon^{eff}
\hat{Q}$. A next step is to prove the convolution formula
\ba
U_{2\epsilon}^{ph}(u,u')&=&\int\limits_Kd^Mu_1\mu(u_1)
U_\epsilon^{ph}(u,u_1)U_\epsilon^{ph}(u_1,u')\\
&=&\int\limits_{\R^M}\frac{d^Mu''}{(\mu\mu'')^{1/2}}
U_{2\epsilon}^{eff}(u,u'')Q(u'',u')\ ,\\
U_{2\epsilon}^{eff}(u,u'')&=&\int\limits_{\R^M}d^Mu''
U_\epsilon^{eff}(u,u'')U_\epsilon^{eff}(u'',u')\ ,
\ea
or in the operator form
\be
\hat{U}_{2\epsilon}^{ph}=\hat{U}_\epsilon^{eff}\hat{Q}\hat{U}_\epsilon
^{eff}\hat{Q} = \hat{U}^{eff}_{2\epsilon}\hat{Q}\ .
\ee
The proof is given by the following chain of equalities
\ba
U_{2\epsilon}^{ph}(u,u')=\sum\limits_{S_F}\int\limits_K d^Mu_1
\frac{\mu_1}{(\mu\mu(\hat{s}u_1))^{1/2}}
U_\epsilon^{eff}(u,\hat{s}u_1)U_\epsilon^{ph}(u_1,u')&=&\\
=\sum\limits_{S_F}\int\limits_Kd^Mu_1|J_s(u_1)|^{1/2}\left(
\frac{\mu_1}{\mu}\right)^{1/2}U_\epsilon^{eff}(u,\hat{s}u_1)
U_\epsilon^{ph}(\hat{s}u_1,u')&=&\\
=\int\limits_{\R^M}
\frac{d^Mu''}{(\mu\mu'')^{1/2}}
\sum\limits_{S_F}
\int\limits_Kd^Mu_1|J_s(u_1)|U^{eff}_\epsilon(u,\hat{s}u_1)
U_\epsilon^{eff}(\hat{s}u_1,u'')Q(u'',u')&,&
\ea
where $\hat{s}u_1=u_s(u_1),\ \mu_1=\mu(u_1),\ \mu= \mu(u)$ and
$\mu''= \mu(u'')$. To obtain (40), we substitute (32) into (36) and
do the integral over $u''$. For the transformation (41) we
use the relation $\mu(u_s) = \mu(u)/J_s(u)$ (which follows from
the $S_F$-invariance of the measure $d^Nx= (\wedge d\theta)d^Mu
\mu(u) = (\wedge d\theta)d^Mu_s\mu(u_s)$) and the $S_F$-invariance
of the kernel (32) (or (30)). The latter results from the
obvious relation
$\langle\hat{s}u|u'\rangle = \langle u|u'\rangle,\ u\in K $
(cf. (31)) and (28). Equality (42) is derived by
substituting (32) into (41) and using the relation $\mu(\hat{s}u_1)
= \mu(u)/J_s(u)$ again. Finally, (42) turns into (37) after
changing variables $u_1\rightarrow \hat{s}u_1$ in each term of the
sum in (42) by means of the rules (9) and (10).

For a finite time we get (see (39))
\ba
\hat{U}_t^{ph}&=& \lim\limits_{\epsilon\rightarrow 0}
\hat{U}_\epsilon^{eff}\hat{Q}\hat{U}_\epsilon^{eff}\hat{Q}\cdots
\hat{U}_\epsilon^{eff}\hat{Q}=\lim\limits_{\epsilon\rightarrow 0}
\hat{U}_\epsilon^{eff}\cdots \hat{U}^{eff}_\epsilon\hat{Q} =
\hat{U}_t^{eff}\hat{Q}\ ,\ \ \ \ \ \\
U_t^{eff}(u,u'')&=&\lim\limits_{\epsilon\rightarrow 0}\!
\int\limits_{\R^M}\!\!\left(\prod\limits_{k=1}^nd^Mu_k\right)
U_\epsilon^{eff}(u,u_n)U_\epsilon^{eff}(u_n,u_{n-1})\cdots
U_\epsilon^{eff}(u_1,u'')\ .
\ea
Formulas (43-44) and (33) with $\Delta_k/\epsilon = \dot{u}_k
+O(\epsilon)$ solve the problem of the PI construction. Equalities
(44) and (33) determine a standard slice approximation of PI
over an Euclidean phase space. Removing the slice regularization
in (44) we obtain the path integral
\be
U_t^{eff}(u,u'')= \int\limits_{\R^{2M}}
\left(\prod\limits_{\tau = 0}^t\frac{d^Mp(\tau)d^Mu(\tau)}
{(2\pi)^M}\right)\exp i\int\limits_0^td\tau \left(p_j\dot{u}_j
-H^{eff}\right)\ ,
\ee
where the measure implies a sum over all trajectories $u(\tau)$
going from the initial point $u''=u(0)$ to the final one $u=u(t)$.
The physical transition amplitude is given by (32) ($\epsilon
\rightarrow t$) and implies a sum over trajectories going from
a {\em few} initial points $u_s(u')= \hat{s}u', \ u'=u(0),\
\hat{s}\in S_F$, to the final one.
A trajectory going from one of these points, say,
$\hat{s}u'\in K_s,\ \hat{s}\neq 1$, to $u=u(t)\in K$ must cross
the boundary $\pl K$ at a point $\tilde{u}=u(\tilde{\tau})$.
Suppose for simplicity that $u(\tau)\in K$ if $\tau\in (\tilde
{\tau},t)$ and $u(\tau)\in K_s$ if $\tau\in (0,\tilde{\tau})$.
Consider a reflected trajectory composed of two pieces
$\hat{s}^{-1}u(\tau),\ \tau\in (0,\tilde{\tau})$ and $u(\tau),\
\tau\in (\tilde{\tau},t)$, i.e. it connects the initial point
$u'\in K$, the "reflection" point $\tilde{u}\in \pl K$ and the
final point $u\in K$. Due to the $S_F$-invariance of the
effective action, the reflected trajectory gives the same
contribution into the sum over pathes as the "straight" one
$u_s(u')\rightarrow u$. {\em Therefore the PI modification
(43) due to a non-trivial topology of $\ph$ (or $CS_{ph}$)
means that in addition to "straight" trajectories $u'\rightarrow u$,
the reflected trajectories $u'\rightarrow \pl K \rightarrow u$
must be included into the sum over paths}.

Contributions of the "reflected" trajectories modify a semiclassical
approximation of PI \cite{book}. For instance, in minisuperspace
cosmological models with gauge fields \cite{ze}, the Hartley-Howking
wave function of the universe depends on the structure of a physical
configuration (phase) space of gauge fields \cite{cos}.

An analysis of quantum Green functions on the gauge orbit space
can be found in \cite{gr}.

Applications to gauge field theories is given in \cite{gr},\cite{pisa}.
In the case of 2D Yang-Mills theories ($QCD_2$), the path integral (45)
can be explicitly done, which allows us to verify the method of reflected
trajectories (formula (32)) by comparing its results with the known ones
obtained in the framework of the loop approach \cite{mig},\cite{raj}.
For $QCD_2$, $K$ is the Weyl cell (a polyhedron), $S_F$ coincides with
the affine Weyl group \cite{pisa}. Contributions of trajectories
reflected from the boundary $\pl K$ (which is provided by the operator
$\hat{Q}$ entering into the PI formula) are shown to be necessary to recover
results of the loop approach \cite{pisa}.

The method can be also generalized to gauge models with fermions
(with anticommuting (Grassmann) variables) \cite{jpa},\cite{book}.
It is curious that the operator $\hat{Q}$ depends on Grassmann variables,
which is important for a correct calculation of a semiclassical
transition amplitude in minisuperspace cosmological gauge  models
with fermions \cite{ibe}.

\end{document}